\UseRawInputEncoding
\documentclass{vgtc}                          % final (conference style)
\graphicspath{{figures/}{pictures/}{images/}{./}} % where to search for the images

\usepackage{times}                     % we use Times as the main font
\usepackage{tabu}                      % only used for the table example
\usepackage{booktabs}                  % only used for the table example
\usepackage{lipsum}                    % used to generate placeholder text
\usepackage{mwe}                       % used to generate placeholder figures

\usepackage{mathptmx}                  % use matching math font

\onlineid{0}

\vgtccategory{Research}

\vgtcinsertpkg

\title{Identifying Framing Practices in Visualization Design Through Practitioner Reflections}

\author{Prakash Shukla\thanks{e-mail: shukla37@purdue.edu}\\ %
        \scriptsize Purdue University %
\and Paul C. Parsons\thanks{e-mail: parsonsp@purdue.edu}\\ %
     \scriptsize Purdue University %
}

\abstract{
    Framing---how designers define and reinterpret problems, shape narratives, and guide audience understanding---is central to design practice. Yet in visualization research, framing has been examined mostly through its rhetorical and perceptual effects on audiences, leaving its role in the design process underexplored. This study addresses that gap by analyzing publicly available podcasts and book chapters in which over 80 professional visualization designers reflect on their work. We find that framing is a pervasive, iterative activity, evident in scoping problems, interpreting data, aligning with stakeholder goals, and shaping narrative direction. Our analysis identifies the conditions that trigger reframing and the strategies practitioners use to navigate uncertainty and guide design. These findings position framing as a core dimension of visualization practice and underscore the need for research and education to support the interpretive and strategic judgment that practitioners exercise throughout the design process.
} % end of abstract


\begin{document}


\firstsection{Introduction}

\maketitle
% An important aspect of data visualization is to communicate, wherther it's insights from data or a narrative around it. The way in which it is communicated can have significant impact on the users, often called framing effect. Researchers including Hullman etc, have studied ... However, not much have been explored about framing practices of designers, 

Framing is a powerful mechanism for shaping how people interpret information. In communication theory, framing refers to the selection and salience of specific aspects of a complex reality to promote a particular interpretation \cite{entman_framing_1993}. It has been studied across disciplines such as communication \cite{entman_framing_1993}, sociology \cite{goffman_frame_1974}, linguistics \cite{fillmore_frame_2006}, and decision--making \cite{tversky_framing_1981}, where it is known to influence how audiences perceive causes, assign responsibility, and evaluate choices. In data visualization research, framing has often been discussed in terms of its rhetorical and perceptual effects on viewers---for example, how editorial choices in annotation, data selection, or interactivity shape what an audience sees and how they interpret it \cite{hullman_visualization_2011}. While important, this audience-centered view overlooks an equally critical dimension: how framing unfolds during the design process itself. Without an account of framing in design cognition, visualization research risks underrepresenting the interpretive judgment that practitioners bring to their work. 

In design theory, framing refers to how designers define and redefine problems, interpret constraints, and construct narratives that make sense of ill-structured situations. Donald Sch\"{o}n \cite{schon_reflective_1983} described designers as engaging in a \textit{reflective conversation} with the situation, in which problems are not merely solved, but framed and reframed in response to evolving understandings. From this perspective, designers are not merely problem solvers but also problem framers---actively shaping how issues are defined, what counts as data, and which interpretations become possible. Yet we know little about how visualization practitioners themselves describe this activity.

This paper addresses that gap by examining framing from the perspective of professional visualization designers. We analyze a corpus of publicly available podcasts and book chapters in which over 80 designers reflect on their work. These narratives offer a unique window into framing activity across the design process—how practitioners scope problems, interpret data, align with or resist stakeholder goals, and anticipate audience reactions. Our analysis surfaces the cognitive and judgmental work that underlies communicative visualization and advances three core arguments: (1) framing is central to professional practice, shaping how problems are defined, what data are emphasized, and which narratives are constructed; (2) framing is best understood as a form of design judgment—an interpretive and strategic capacity enacted iteratively across scoping, data engagement, narrative, and visual design; and (3) recognizing framing as judgment has implications for visualization research and education, challenging process models that underrepresent interpretive work and underscoring the need to teach framing as a core professional skill rather than a by-product of technical proficiency.

\section{Framing in Visualization Design: From Rhetoric to Practice}
Framing has long been recognized as a mechanism through which communicators shape interpretation. It underlies how media influence public discourse \cite{lakoff_metaphors_1980}, how choices are presented in decision–making \cite{tversky_framing_1981}, and how people construct meaning in social interaction \cite{goffman_frame_1974}. In this tradition, framing is studied primarily for its effects on audiences: how the presentation of information influences interpretation, emotion, and behavior.

Visualization research has followed this audience-centered orientation, examining how rhetorical choices in visualizations shape perception. Scholars have argued that effective visualization requires attention not only to encoding but to how meaning is interpreted and communicated \cite{franconeri_science_2021}. Designers are cast as narrative architects: Kosara and Mackinlay \cite{kosara_storytelling_2013} emphasized storytelling as the future of visualization, while Hullman and Diakopoulos \cite{hullman_visualization_2011} demonstrated how data selection, visual emphasis, and annotation guide interpretation.

These works highlight visualization’s rhetorical power but leave underexplored the designer’s own framing activity---how practitioners define, interpret, and reframe problems as part of the design process. In design theory, framing refers to the cognitive and reflective acts through which designers make sense of ambiguous, ill-structured situations. Sch\"{o}n \cite{schon_reflective_1983} described design as a reflective conversation with the situation, where problems are continuously reframed in response to new constraints and insights. Dorst \cite{dorst_co-evolution_2019} extended this view with the idea of problem–solution co-evolution, while recent studies show that visualization practitioners also navigate these dynamics \cite{parsons_beyond_2025}. Together, this literature points to framing as a central mechanism of design cognition---yet its role in visualization practice remains empirically underexamined.

Our study addresses this gap by analyzing podcasts and book chapters featuring over 80 professional visualization designers. By examining how practitioners describe their own work, we investigate how framing informs problem definition, design decisions, and communication strategies. In doing so, we shift the conversation from framing as audience effect to framing as design judgment, surfacing the interpretive labor that often remains implicit in accounts of visualization outcomes.

\section{Methods}
To investigate whether and how framing occurs in data visualization practice, we analyzed publicly available content in which practitioners reflect on their design processes---specifically podcasts, book chapters, and online videos. This approach provides access to spontaneous, first--person accounts of design reasoning \cite{turner_podcast_2023}, using a \textit{podcast ethnography} method \cite{lundstrom_podcast_2021} that captures naturally occurring, unsolicited reflections and conversational insights into how practitioners describe their decisions. While unscripted content presents credibility and consistency challenges \cite{drew_educational_2017}, it offers rich perspectives from experienced industry professionals \cite{noauthor_podcasts_nodate, noauthor_storytelling_nodate, braun_data_2017}. This method is well--suited for identifying whether framing occurs in practice and offers preliminary insight into its nature, providing broad access to diverse practitioners across visualization domains.

The dataset includes 23 episodes from the \textit{Storytelling with Data} podcast (average length: 50 minutes) \cite{noauthor_storytelling_nodate}, 22 \textit{PolicyViz} video podcasts (average length: 40 minutes) \cite{noauthor_podcasts_nodate}, and 40 expert interviews from the book Data Visualization for Success (average length: 450 words) \cite{braun_data_2017}. Quotes from these sources are labeled as S1, S2..., P1, P2..., and B1, B2..., respectively. These sources included responses to questions on design methods, elements, applications, and advice. Interviews with individuals not engaged in professional design practice, such as those in academia, were not included.

Transcripts were analyzed using hybrid thematic analysis in Dovetail by the first author. Deductive codes, drawn from the framing literature, included cognitive framing (interpretive moves), semantic framing (language/terminology choices), and communicative framing (shaping audience meaning) \cite{prendeville_politics_2022}, as well as design markers such as \textit{novel standpoint} \cite{dorst_frame_2015}, \textit{active perspective} \cite{kolko_sensemaking_2010}, and \textit{selection and salience} \cite{entman_framing_1993}. To enhance data familiarization, podcasts were reviewed and book chapters were read multiple times. Special attention was given to instances where practitioners justified their choices, adjusted designs based on audience or context, or articulated their rationale. While reviewing the transcripts, new codes were captured that were not covered in the theory (e.g., communicating uncertainty, lived experience, and getting the ``right story'')

The analysis examined: (1) when and where framing occurs, (2) what conditions prompt it, and (3) strategies practitioners use to frame problems.

\section{Findings: Evidence of Framing as Design Judgment}

\subsection{Framing across Design Activities}

Framing in visualization practice does not follow a linear sequence but instead emerges across a range of design activities, often in iterative, overlapping, and recursive ways. Rather than treating framing as a discrete step in a predefined process, practitioners engage in framing as they scope problems, explore data, shape narratives, and construct visual representations. In this section, we examine how framing operates within these activities, drawing on examples that illustrate its situated and evolving role in real--world design work. See \autoref{tab:framing_in_phases} for examples.

\subsubsection{Scoping and Contextualizing the Problem} 
Practitioners frequently engage in framing when interpreting the scope and intent of a project. This includes identifying objectives, aligning with audience needs, and determining how best to position the data.

S1 captures the foundational role of framing early in the process: \textit{``figuring out what the data is about and what you should chart, how you should chart it, and why you should do it''} makes the rest of the process easier. This kind of interpretive framing sets the direction for subsequent design decisions. Often, it involves questioning the initial brief to uncover the real problem. As P5 explains, \textit{``instead of emailing back a spreadsheet, we'll call them and ask, 'What are you trying to do?'''} This act of inquiry reveals how practitioners rearticulate design goals in relation to broader project intentions and constraints. Importantly, this scoping is not done in isolation---it is often shaped collaboratively with stakeholders. As P1 notes, \textit{``most of the time, I'm together with the people shaping the story themselves.''} Rather than accepting the problem as given, practitioners co--construct its contours in dialogue with stakeholders, illustrating how framing is embedded in the social dynamics of design work.

\begin{table*}
  \centering
  \caption{Examples of framing across key design activities in data visualization practice}
  \label{tab:framing_in_phases}
\begin{tabular}{p{4.2cm}p{5.2cm}p{6.8cm}}
    \toprule
    Design Activity & Framing Focus & Illustrative Practices \\
    \midrule
    Scoping and Contextualizing the Problem & Defining purpose, identifying audience, interpreting goals & Clarifying objectives, uncovering implicit needs, negotiating scope \\
    
    Engaging with Data & Selecting, filtering, and interpreting data & Choosing representative samples, valuing granularity, simplifying complexity \\
    
    Constructing a Narrative & Shaping the story the data tells & Highlighting the unexpected, connecting data to human experience \\
    
    Designing Visuals to Guide Interpretation & Selecting encodings and emphasis strategies & Using layout, color, titles, and annotations to focus attention \\
    \bottomrule
\end{tabular}
\end{table*}

\subsubsection{Engaging with Data} 
Another area where framing occurs is while engaging with data through selection, analysis, and interpretation. This phase centers on identifying what matters, filtering out noise, and building the foundation for a meaningful narrative. P2 highlights the power of simplification: \textit{``You don't need big data... take out small sections... to represent the bigger picture. Through narrowing down, you often gain a new understanding.''} Similarly, S3 stresses the importance of tailoring detail to the audience: \textit{``It's not always about telling the two--hour version... sometimes you need just a couple of sentences.''} Further, framing in this phase can also involve leaning into granular data, as S1 highlights: \textit{``I look for the highest granularity dataset I can find, even if others might see it as noise. A lot of the time, that noise is a better reflection of reality, so I try to show as much of it as I can while still revealing the pattern.''} These choices---what to emphasize, what to exclude---ultimately shape how insights are constructed and communicated.

\subsubsection{Crafting a Narrative} 
Framing in this phase involves more than simply presenting data---it is about shaping a perspective and telling a compelling story. This narrative framing can take multiple forms, such as revealing the unexpected, adding a human dimension to otherwise abstract information, or persuading the audience. B6 emphasizes the importance of uncovering the unexpected in routine datasets: \textit{``most data you look at won't surprise you, so I always try to look for the unexpected and the compelling hidden inside fields of boring data.''} This pursuit of surprise transforms visualizations from routine to revealing. Similarly, B1 reflects on the value of visual storytelling in surfacing the unseen: \textit{``Our main goal is always to show a new aspect of the data or make something visible that wasn't visible before. As John W. Tukey said, ‘The greatest value of a picture is when it forces us to notice what we never expected to see.'''} Alongside this pursuit of insight, narrative also brings a human element to the data. As P4 puts it, \textit{``it doesn't just show you the hard facts---it shows how those hard facts reflect on the lives being represented''}. This type of framing invites empathy and grounds abstract information in real--world context. Moreover, framing can serve a persuasive function. As S10 explains, \textit{``We are all trying to glean insight quickly and often persuade people to see things they didn't see before. If you do data visualization right, you're helping them process information more effectively and find insight faster.''} In this sense, framing is not just about what is shown, but how it guides interpretation and meaning.

\subsubsection{Designing Visuals to Guide Interpretation} 
Practitioners use visuals to tell a specific story. This involves not only selecting appropriate visual forms (e.g., charts, graphs, or maps) but also making intentional design choices, such as layout, color, typography, and interaction, to shape how audiences interpret the data. For example, P12 highlights how color influences meaning in a visualization: \textit{``The blue versus red is really striking, not only for drawing your attention to those areas in red, and kind of that signaling of when we do see red, we tend to think of some sort of calamity going on in some ways.''} Similarly, S12 notes how even subtle visual tweaks can shift perception, explaining how choropleth maps behave differently on light versus dark backgrounds: \textit{``Darker means more unless you're on a dark background, then the inverse works better.''} Text also plays a role in framing interpretation. S2 shares how annotating a chart altered its message: \textit{``I created a line chart showing how avocado prices changed over time, but I thought, let's find a different angle. So, instead of focusing on the price differences, I highlighted how now is the best time to make guacamole because organic avocado prices are at their lowest.''} In this case, the same dataset---presented with a new narrative---became more engaging and relevant to the audience.

\subsection{Factors Influencing Framing Judgments}
Although not exhaustive, in our analysis, we have noted several factors that create a situation in which designers are more likely to make framing judgments.

\subsubsection{Taming Complexity} 
Framing judgments are often prompted by complexity--when designers must make sense of overwhelming, ambiguous, or unwieldy data. Large datasets, for instance, require framing to identify priorities and focus the design. As P5 notes, \textit{``we have thousands of indicators... and ended up helping shape the questions they're trying to answer''} Here, framing helps reduce informational overload and guide insight. Framing is also triggered by the challenges of working with unstructured or high--cardinality data. P27 describes the limits of standard tools: \textit{``You run into problems... What happens when you have more than 10 categories?''} In these cases, framing helps designers choose appropriate representations and avoid oversimplification. Even small datasets can prompt framing due to the vast number of potential interpretations. S1 explains, \textit{``There are thousands of ways to visualize something... you want to filter it down to highlight what matters.''} In each case, framing emerges as a response to complexity, helping designers structure data, define focus, and select visuals that support meaningful interpretation.

\subsubsection{Designers' Personal Touch} 
Designer's personality, perspective, and data sensibilities also influence framing. B3 explains, \textit{``Once I have an opinion about what's interesting, I decide how to move it from my head to the audience so others can share my perspective.''} Some, like B2, focus on finding the right dataset: \textit{``I'm more concerned about finding the right dataset than the right representation.''} Others, like S1, embrace data ``noise'' as a reflection of reality, and B18 emphasizes aesthetic impact: \textit{``We try to make sure the viewer doesn't fall asleep.''} Ultimately, designers bring their own values, preferences, and goals to the framing process, contributing to the diversity and richness of visualization styles.

\subsubsection{Adapting to Diverse Objectives: Framing for Different Purposes} 
Framing is shaped by the purpose of the visualization, whether it's for exploration, explanation, or narration. As S4 notes, \textit{``Data visualization can have a multiplicity of purposes.''} Exploratory views may show full data distributions, while explanatory ones highlight specific trends. The nature of the audience also influences framing choices. S3 describes an \textit{``empathy spectrum''} where designers tailor complexity and detail to fit viewer needs. B5 adds that framing can evoke emotional resonance: \textit{``I look for the fundamentally human angle... so people find something about themselves reflected in it.``} These varying goals and audiences prompt designers to frame data differently, guiding what they emphasize, simplify, or humanize.

\subsection{Framing Strategies}
Practitioners employ a range of framing strategies throughout the design process. They employ a process of inquiry and exploration. Many begin by asking critical questions about the data, audience, and objectives to clarify intent. P14 described starting with a \textit{``discovery report''} to define what the data is about, who the audience is and what it should convey. Some, including P18, highlight the need to converse with data collectors to verify accuracy: \textit{``An outlier doesn't have to be an outlier, it can be a typo.''} Beyond questioning, practitioners immerse themselves in the data to discover visual patterns. B11 describes using sketching to explore possibilities: \textit{``These sketches are often messy and unstructured,''} while P3 emphasizes analysis through tools like R: \textit{``I pull the data into R and then explore for structures and visualization methods.''}

Collaboration and iteration are equally central to framing. Practitioners often refine the problem space and narrative through dialogue with stakeholders. S7 describes working closely with users to understand interpretation: \textit{``We talked to some users about how they read the chart.''} S4 notes how iteration helps surface the right message: \textit{``We did a lot of attempts... it allowed us to talk with the client to slowly get to what would be the message.''} This co--construction process ensures alignment with the client's intent and audience expectations. S15 reinforces this, stating, \textit{```If it doesn't work for the end user, then there's not much sense in building your visualization.''} By centering users in the framing process, practitioners make design decisions that are both functional and meaningful.

Finally, practitioners shape framing through both visual and textual emphasis. P11 describes minimizing noise to sharpen the message: \textit{``I am showing you in a specific restricted context what matters, and I am taking down what doesn't matter.''} Text also plays a vital role in guiding interpretation. S6 refers to text as ``a first--class citizen'' in visualizations, helping explain causes and highlight relevance. Together, these strategies ensure the story told through the data is intentional, accessible, and impactful.

\section{Discussion}
\subsection{Framing as Situated and Layered Judgment}
In our findings, we observe that framing in data visualization practice is not limited to a single early--stage activity such as project scoping or goal setting. Rather, framing is manifested across multiple levels throughout the design process. These include how practitioners define the scope of the project, engage with data, construct narratives, and design visual representations. This idea that framing occurs throughout the design process has also been noted in broader design related fields. For example, Nelson and Stolterman \cite{nelson_design_2002} describe framing as a key cognitive act for navigating design complexity. Our analysis aligns with these views, revealing that framing occurs at multiple levels in data visualization design.

Importantly, we also note that these levels are not sequential phases but interwoven and mutually informing. Designers don't just frame the problem and then move on to data or visuals in a linear way; rather, this process involves cycles of reflection and action. For example, S2 initially created a standard price comparison chart but later reworked the narrative by asking, \textit{``why would this data be interesting to someone at this particular time?''} This shift---from describing trends to highlighting that \textit{``now is the time to make guacamole''}---transformed how the same data was visually and rhetorically expressed. This illustrates how insight at one layer---e.g., a visually displayed chart---can reshape framing other aspects of the situation like the narrative. This finding exemplifies the interwoven and co--evolutionary nature of design practice that has been articulated by researchers studying design practice \cite{schon_reflective_1983, dorst_co-evolution_2019, parsons_beyond_2025}. Our findings resonate with recent work that conceptualizes judgment in visualization design not as an isolated act but as a form of distributed coordination across people, tools, and systems \cite{parsons_judgment_2025}. This systems-level perspective complements our analysis of framing by foregrounding how interpretive moves are sustained through alignment, adaptation, and negotiation in practice.

Because of this intertwined structure, being effective as a data visualization designer requires making sound judgments for fluid movement between framing levels. Designers essentially have to wear multiple hats, including the role of analyst, storyteller, and visual communicator, making the practice of visualization design both science and art \cite{schwabish_practice_2021}. This demands that designers engage in continuous interpretation and reflective judgment--shifting between roles of analyst, storyteller, and visual rhetorician--as they navigate evolving constraints and opportunities.

\subsection{Framing as a Pedagogical Target}
Given the pervasive role of framing in expert practice and its situated, layered nature, there is a clear need to develop effective methods for teaching these skills to novices and students. We argue that framing should be taught as an interpretive skill rather than a prescriptive procedure. In practice, this means helping students learn to actively interpret open-ended design situations and make reflective judgments about how to define and approach a visualization problem, instead of following rules or procedures. Research in design and engineering education reinforces this need: framing is often the most difficult yet important part of tackling \textit{wicked problems}--complex challenges with no single right answer \cite{beckman_teaching_2012}.

Research on teaching framing skills is limited, although some ideas have been proposed in adjacent areas like UX design \cite{parsons_developing_2023}. These strategies include enabling students to work on multiple short, end--to--end design projects and intentionally incorporating open ended project prompts. End to end projects help students practice framing continuously throughout the design process rather than only at the outset. Working on multiple projects allows students to build a repertoire of design knowledge, which they can leverage to refine their framing skills. Open--ended projects, in particular, require students to define the project scope, a core aspect of framing. There are various techniques for forcing students to frame and re-frame during their project work. For instance, injecting failure into students' experiences can help students articulate risks, recognize blind spots, and improve the quality of their planning and collaboration---e.g., using a premortem technique \cite{parsons_teaching_2025}. Other fields like management and engineering have also explored strategies for teaching framing \cite{beckman_teaching_2012, svihla_funds_2022}. Beckman and Barry \cite{beckman_teaching_2012} use a storytelling approach, encouraging students to empathize with customer narratives, synthesize diverse perspectives, and iteratively refine the scope based on contextual insights. Alternatively, Svihla et al. \cite{svihla_funds_2022} use a funds of knowledge framework, allowing students to draw on personal and cultural experiences to tackle sociotechnical challenges.

Developing framing expertise requires meta--cognitive awareness \cite{kurt_improving_2017}--the ability to recognize when and how one is engaging in framing and to critically reflect on the choices involved. A studio-like environment that infuses critique and feedback into the experience can contribute to the development of student's framing ability. Parsons et al. \cite{parsons_developing_2023} note that frequent navigational feedback from instructors (asking questions like ``why are you focusing on this issue? Have you considered this perspective?'') often led students to reflect, reconsider and adjust their problem frames. Further, they report that peer critique sessions allowed students to witness how others approached the same open--ended brief with different framings, thus revealing the multitude of valid ways one might define a problem. Treating framing as a situated, debatable, and revisable act helps students develop not just visualization skills, but the critical capacity to shape meaning responsibly in public communication. This study contributes to those efforts by surfacing the types of framing moves and strategies that experienced practitioners describe using in real-world contexts. By articulating these practices in designers' own terms, we offer an empirical foundation for teaching framing not as an abstract concept, but as a situated and learnable form of design judgment.

\section{Limitations and Future  Work}
This study draws on publicly available sources---podcasts and book chapters---where data visualization practitioners reflect on their design processes. While valuable, this approach has limitations: the researchers had no control over data collection, the content is self--reported rather than observational, and time lags may introduce recall bias. As a result, the corpus may emphasize reflective, rationalized accounts of framing rather than tacit or emergent practices. This means our findings likely understate the role of coordination breakdowns and implicit negotiation. These factors may also have shaped the themes that emerged. Although the data revealed framing practices and offered early insight into framing strategies, it did not allow for deeper exploration of subconscious framing judgments. While our corpus revealed explicit practitioner accounts of framing, it was less able to capture the tacit and collective dynamics that shape how framing unfolds in teams. Complementary work on judgment as coordination \cite{parsons_judgment_2025} highlights these invisible dimensions, offering concepts such as coordination repair and sacrifice judgments that future studies might apply to observational or collaborative settings. Despite these limitations, the study provides evidence of the role of framing throughout the design process in visualization practice. While framing was clearly observed, distinctions between conscious and subconscious processes remain unclear. Future research using methods like observations, interviews, diary studies, or design probes may yield richer insights into this complex phenomenon.

\section{Conclusion}
This study highlights the critical role of framing in data visualization design, showing how practitioners continually define problems, interpret data, and shape narratives through iterative judgment. By analyzing podcasts and book chapters, we provide empirical evidence of how framing operates across the design process and identify strategies that practitioners use to navigate uncertainty and guide interpretation. Our study contributes to a growing reorientation in visualization research: seeing design not merely as a sequence of technical choices but as interpretive work shaped through framing. We have shown, first, that framing is a central and continuous element of practice; second, that it operates as design judgment across multiple layers of activity; and third, that acknowledging framing has implications for both research models and educational approaches. By centering framing in accounts of practice, we surface the interpretive labor that underlies visualization design and argue for its recognition as a foundational dimension of the field.

%Our findings underscore that framing is not an afterthought or rhetorical add-on, but a central mode of reasoning through which practitioners define problems, make meaning, and shape audience understanding. Because framing emerges through inquiry, interpretation, and negotiation in practice, we recommend teaching framing not as a fixed step but through open-ended, iterative projects that mirror the ambiguity and situated nature of professional design work. By foregrounding framing as a form of professional judgment, this work contributes to ongoing exploration of how visualization designers think, make meaning, and communicate visually.
%

\acknowledgments{
This work was supported by NSF grant \#2146228.}

\bibliographystyle{abbrv-doi}

\bibliography{references}
\end{document}